\documentclass[sigconf]{acmart}
\usepackage{amsmath}
\usepackage{amssymb}
\usepackage{graphicx}
\usepackage{algorithm2e}
\usepackage{breqn}
\usepackage{multirow}
\usepackage{capt-of}
\usepackage[T1]{fontenc}
\usepackage{letltxmacro}
\usepackage{listings}
\usepackage{color}

\usepackage{subcaption}

\usepackage{bookmark}
\usepackage{tcolorbox}

\definecolor{dkgreen}{rgb}{0,0.6,0}
\definecolor{gray}{rgb}{0.5,0.5,0.5}
\definecolor{mauve}{rgb}{0.58,0,0.82}

\renewcommand\footnotetextcopyrightpermission[1]{} % removes footnote with conference information in first column

\makeatletter
\usepackage{amssymb}
\newcommand{\ygg@basicalert}[2]{\fbox{\bfseries\sffamily\scriptsize#1}{\sf\small$\blacktriangleright$\textit{#2}$\blacktriangleleft$}}

\title{Static Code Analysis of Multilanguage Software Systems}

\subtitle{The Case of Multilanguage Java EE Applications}

\author{Anas Shatnawi, Hafedh Mili, Manel Abdellatif, Yann-Ga\"el Gu\'eh\'eneuc, Naouel Moha, Geoffrey Hecht, Ghizlane El Boussaidi, Jean Privat}
\affiliation{%
  \institution{LATECE Laboratory, Universit\'e du Qu\'ebec \`a Montr\'eal, Canada}
  \city{Montreal} 
  \country{Canada}
}

\begin{document}

\thispagestyle{empty}
\pagestyle{empty}

\begin{abstract}
Identifying dependency call graphs of multilanguage software systems using static code analysis is challenging. The different languages used in developing today's systems often have different lexical, syntactical, and semantic rules that make thorough analysis difficult. Also, they offer different modularization and dependency mechanisms, both within and between components. Finally, they promote and--or require varieties of frameworks offering different sets of services, which introduce hidden dependencies, invisible with current static code analysis approaches. 

In this paper, we identify five important challenges that static code analysis must overcome with multilanguage systems and we propose requirements to handle them. Then, we present solutions of these requirements to handle JEE applications, which combine server-side Java source code with a number of client-side Web dialects (e.g., JSP, JSF) while relying on frameworks (e.g., Web and EJB containers) that create hidden dependencies. Finally, we evaluate our implementations of the solutions by developing a set of tools to analyze JEE applications to build a dependency call graph and by applying these tools on two sample JEE applications. Our evaluation shows that our tools can solve the identified challenges and improve the recall in the identification of multilanguage dependencies compared to standard JEE static code analysis and, thus, indirectly that the proposed requirements are useful to build multilanguage static code analysis.
\end{abstract}

\keywords{Static code analysis, multilanguage software, program comprehension,  dependency call graph, KDM, container services, JEE applications}

\maketitle

\section{INTRODUCTION}
\label{sec:intro}
 
Popular websites such as Google, Facebook, YouTube, and common software systems, such as Mozilla Firefox, Microsoft Word, are now built in heterogeneous multilanguage environments; using heterogeneous components developed following multiple programing languages, like C, C++, Java, JSP, PHP, SQL, and--or XML, with complex interactions \cite{boughanmi2010multi}.

The static code analysis of these multilanguage systems is essential to support software engineers in their tasks, such as program comprehension, optimization, reuse, and service identification, etc. It extracts a \textit{dependency call graph} that represents the dependencies between program elements. 
As we deal with multilanguage systems, we must compute such static dependency call graph in a way that identifies \emph{all} of the dependencies that exist between different program elements, \emph{despite} the different languages, technologies, and runtime environments, which abstract away---and thus obfuscate---such dependencies.

However, identifying such dependency call graph by analyzing statically the code of multilanguage systems is challenging because: 
(1) different languages \textit{follow different ways to describe their lexical, syntactical, and semantic} rules, 
(2) different languages offer \textit{different modularization mechanisms and manage dependencies differently}, both within and between components, 
(3) several \textit{indirect dependencies are described in textual configuration files} with different syntax and whose interpretation depends on the used frameworks,
(4) heterogeneous components can be \textit{distributed} in different layers (e.g., view, service, and data layers) making it difficult to trace cross-component dependencies through various communication mechanisms, and
(5) heterogeneous components \textit{rely on a variety of frameworks} that manage their life cycles and offer different container services that hide dependencies.

\begin{figure*}[h]
\begin{center}
\includegraphics[width=\textwidth]{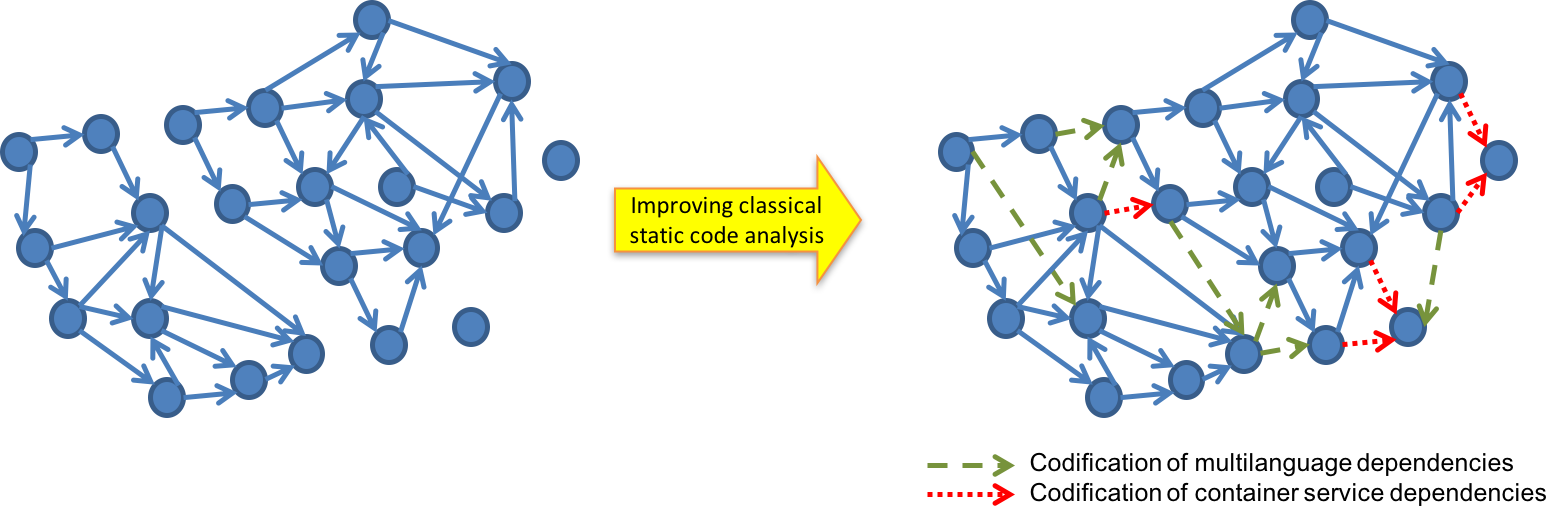}
\caption{Improving classic static code analysis to include multilanguage and container service dependencies}
\label{fig:dependency-call-graphgs}
\end{center}
\end{figure*}

Thus, existing traditional \emph{unilingual} static code analysis approaches must be improved by considering other types of analyses, involving other kinds of software artifacts (e.g., configuration files) and codifying container services provided by frameworks. We illustrate in Figure \ref{fig:dependency-call-graphgs} how the results of a classical static code analysis can be improved with the detection of hidden dependencies by analyzing multilanguage and container service dependencies to connect the different parts of the dependency call graph.

In the literature, approaches have been proposed to support the static code analysis of multilanguage systems, such as \cite{von1995program,kienle2010rigi,bruneliere2014modisco,moise2005extracting,kraft2008cross,german2009license}. Yet, these approaches suffer from three main limitations. First, they can only analyze one programming language at a time, although they can deal with more than one programming language in isolation from one another, like \cite{von1995program,bruneliere2014modisco,kienle2010rigi}. Second, they do not identify hidden dependencies related to container services \cite{bruneliere2014modisco,moise2005extracting,kraft2008cross,german2009license}. Third, they cannot analyze source code files mixing multilanguage code \cite{von1995program,kienle2010rigi,bruneliere2014modisco,moise2005extracting}. 

Moreover, none of these existing approaches described the challenges---and their requirements---of the static analysis of multilanguage software systems. Instead, each focused on providing solutions to a narrow, particular subset of the challenges 1--5.

In this paper, we introduce and discuss the five challenges to analyze statically multilanguage software systems. We identified these challenges based on our literature review, our experience with analyzing JEE applications, and discussions/experience with industrial partners. We propose requirements to handle these challenges and allow the development of static code analyses for multilanguage software systems.

Then, we illustrate how the five challenges can be overcome in the particular case of JEE applications, which are multilanguage software systems that combine server side Java code with a number of client side Web dialects (JSP, JSF, etc.). In JEE applications,  dependencies within and between languages are hidden in container services and various ad-hoc configuration files. 

We develop an automated tool that represents program elements and their related dependencies within and between JEE components as one dependency call graph. To define this dependency call graph, we rely on the OMG's Knowledge Discovery Meta-model (KDM) \cite{perez2011knowledge}, a language-independent meta-model. We also rely on the MoDisco Eclipse plug-in that offers an open-source implementation of KDM for Java code \cite{bruneliere2014modisco}.
We propose a number of rules to transform multilanguage JEE source code to the KDM model and to codify dependencies related to the container services, string literals, and configuration files. We apply our tool on two JEE applications: Java PetStore and JSP Blog. We thus show that our tool can address the five challenges for JEE applications.

The rest of this paper is organized as follows. Section \ref{sec:multilanguageChallenges} discusses the five challenges---and identify their related requirements---for analyzing statically multilanguage software systems to build a dependency call graph. Section \ref{sec.motivation-example} provides a motivating example. Section \ref{sec:change-imact-jee} presents our static analysis approach for JEE applications. Section \ref{sec:evaluation} shows and discusses experimentation results. Section \ref{sec:related-work} presents related works. Finally, Section \ref{sec:conclusion} conclude with future works.

\section{Challenges with Multilanguage Software Systems}
\label{sec:multilanguageChallenges}

We identify five challenges that make static code analyses difficult for multilanguage software systems. These challenges are: (1) diversity of meta-models, (2) mixing multilanguage code in one source code file, (3) various patterns to codify dependencies, (4) hidden dependencies related to frameworks and container services, and (5) string literals to describe dependencies. We analyze these challenges to derive requirements that a static analysis must meet.

\subsection{Diversity of Meta-models}

Different programming languages were designed following different meta-models describe different lexical, syntactic, and semantic rules.
For example, while C++, C\#, Java follow the \textit{object-oriented paradigm}, i.e., their meta-models offer the concepts of \textit{class, method, and attribute}, many other concepts are language specific, e.g., virtual method, copy constructor, interfaces. Further, their elements make communicate based on \textit{method invocation, attribute access}, \textit{inheritance}, \textit{object instantiation}, etc. As other example, ASP, HTML, and JSP follow the \textit{XML meta-models} in their source code files, i.e., their elements are based on \textit{XML tags} that define their attributes and dependencies. Some other programming languages are defined based on the \textit{procedural meta-models}, such as Basic, C, COBOL, etc.

There are situations where one single statement/character has contradicting semantics across different programming languages. For example, the \textit{static} keyword has different meanings in C, C++, and Java. Unlike Java, C++ does not support static blocks.
Therefore, the use of one common code analyzer is difficult.  %\YANN{https://en.wikipedia.org/wiki/Island_grammar} \anas{I comment the whole text, we need to discus this} \YANN{We must discuss Island grammars here and--or in the related work.}

%\begin{tcolorbox}
\textbf{Requirement 1:} the analysis should rely on a standard representation that can represent all programming languages in a multilanguage software system, including program elements and their dependencies.
%\end{tcolorbox}

\subsection{Mixing Multilanguage Code in Source Files}

Some programming languages allow mixing multilanguage code inside a single source code file, e.g., developers are permitted to mix Java and JavaScript code with JSP and HTML tags in JSP technology. Similarly, it is possible to mix HTML, Javascript, and PHP in a single PHP file. The different pieces of code can have internal dependencies among one another inside the same file and external dependencies with code in other files. A static analysis must identify both types of dependencies for completeness.

%\begin{tcolorbox}
\textbf{Requirement 2:} the analysis must detect when multilanguage code is used in one file, parse each piece of code following its language to identify program elements and their internal and external dependencies.
%\end{tcolorbox}

\subsection{Patterns of Dependencies} 

There is a variety of patterns that reflect dependency codification in multilanguage software systems due to: 

\begin{enumerate}
\item \textit{Different communication protocols:} heterogeneous components rely on communication patterns to encode their dependencies. Some components use RMI while some others use HTTP requests, etc. 
Such communication patterns must be captured to codify related dependencies.

\item \textit{Several versions:} programming languages evolve over time. Thus, several mechanisms can be used to codify the same dependencies in every languages. We must identify these mechanisms to capture their dependencies.

\item \textit{Variance in development and design patterns:} developers follow different patterns while developing their systems.
For example, in the \textit{event-driven pattern}, many dependencies are based on runtime events like user actions, sensor values, etc. Other patterns codify dependencies differently\footnote{There are two main challenges behind having different codification patterns. First, developers are constrained to use specific patterns following the application context (e.g., mobile, desktop, Web, etc.). Second, developers have varied background, knowledge, and experience.
Therefore, the analysis must understand how these patterns work to identify related dependencies.}. 

Although different patterns must confirm to the main guidelines and  mechanisms provided by the languages, some languages leave space for developers to make specific patterns of dependencies that cannot be recovered by general static code analysis. Such cases can be found frequently in JEE custom tag libraries where developers have flexibility in implementing tag handler classes\footnote{https://docs.oracle.com/cd/E17802_01/blueprints/blueprints/code/jps11/src/ com/sun/j2ee/blueprints/petstore/taglib/list/PrevFormTag.java.html}.
\end{enumerate}

%\begin{tcolorbox}
\textbf{Requirement 3:} the analysis must know the communication patterns available in each programming language of interest. It must also consider the versions of programming languages. 
%\end{tcolorbox}

\subsection{Hidden Dependencies with Frameworks and Containers}

\begin{figure}[ht]
\centering
\includegraphics[width= 0.5\textwidth]{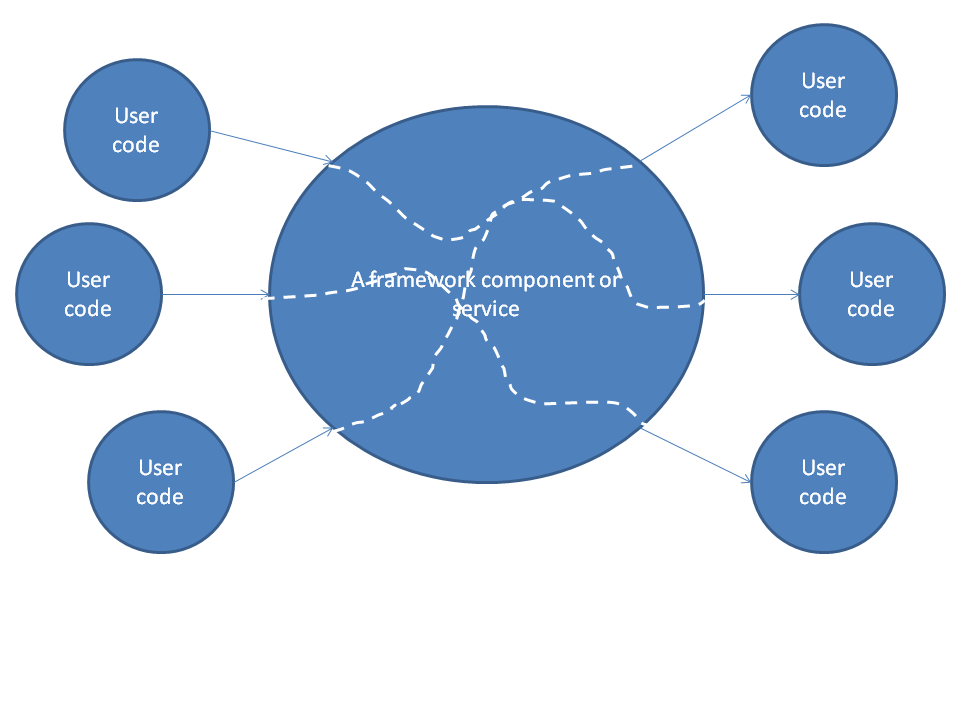}
\caption{Example of hidden dependencies inherent in frameworks}
\label{fig.dealing-with-frameworks}
\end{figure}

Various programming languages rely on different frameworks to deploy their components. These frameworks provide callback methods that \textit{are invoked by the frameworks} at specific application execution times/events. Thus, we must understand these frameworks to identify which callback methods are called at certain times/for certain events, e.g., when the user presses a button. Also, we must explicitly codify/represent the dependencies that are managed by the frameworks. Figure \ref{fig.dealing-with-frameworks} illustrates the idea of hidden dependencies being invisible in user code. It shows that hidden dependencies are caused by:

\begin{itemize}
\item Calls related to the application life-cycle management. For example, when we create a Servlet, the server container automatically calls the \textit{init()} method, then the \textit{service()} one, which can in turn call other methods thus creating hidden dependencies.

\item Configurations files describing applications/frameworks functionalities. These files are located in different directories, which requires visiting these directories to parse each file and identify dependencies.

\item Callback methods offered by frameworks. These methods are used by developers to access underlying services and create hidden dependencies.
\end{itemize}

%\begin{tcolorbox}
\textbf{Requirement 4:} we must analyze the different frameworks used to deploy different components so that the analysis can detect related hidden dependencies. We must understand the life cycle of each component, how they are configured in configuration files, and what callback methods they use to communicate with other components/frameworks.
%\end{tcolorbox}

%\subsection{Reliance on configuration files}
%Many indirect dependencies between components are described using configuration files that have no standard syntax. We need to identify these configuration files, and parse each one following the application domain and tools consuming it. Also, configuration files are located at different directories which requires to visit these directories to interpret dependencies.

%\begin{tcolorbox}
%\textbf{Requirement 5:} the analyzer should be award of what  configuration files, where they are located, how they are interpreted, and what information they carry.
%\end{tcolorbox}

\subsection{String Literals Encoding Dependencies}

String literals are frequently used to realize dependencies and to encapsulate parameter values, e.g., dependencies described in MANIFEST.MF files. The analysis of such literals is difficult because the meaning of a literal depends on the application, frameworks, and programming languages. It may require data-flow analysis to extract the value of the literals. 
%For example, considering the \textit{(id,value)} pattern, the \textit{id} should be compatible in the source and the destination, otherwise it brings \textit{null} values.

%\begin{tcolorbox}
\textbf{Requirement 5:} the analysis must identify where/when string literals are used to encode dependencies and know how to parse each string literal following the application context. 
%\end{tcolorbox}

\section{Motivating Example}
\label{sec.motivation-example}

\begin{figure*}[ht]
\begin{center}
\includegraphics[width=\textwidth]{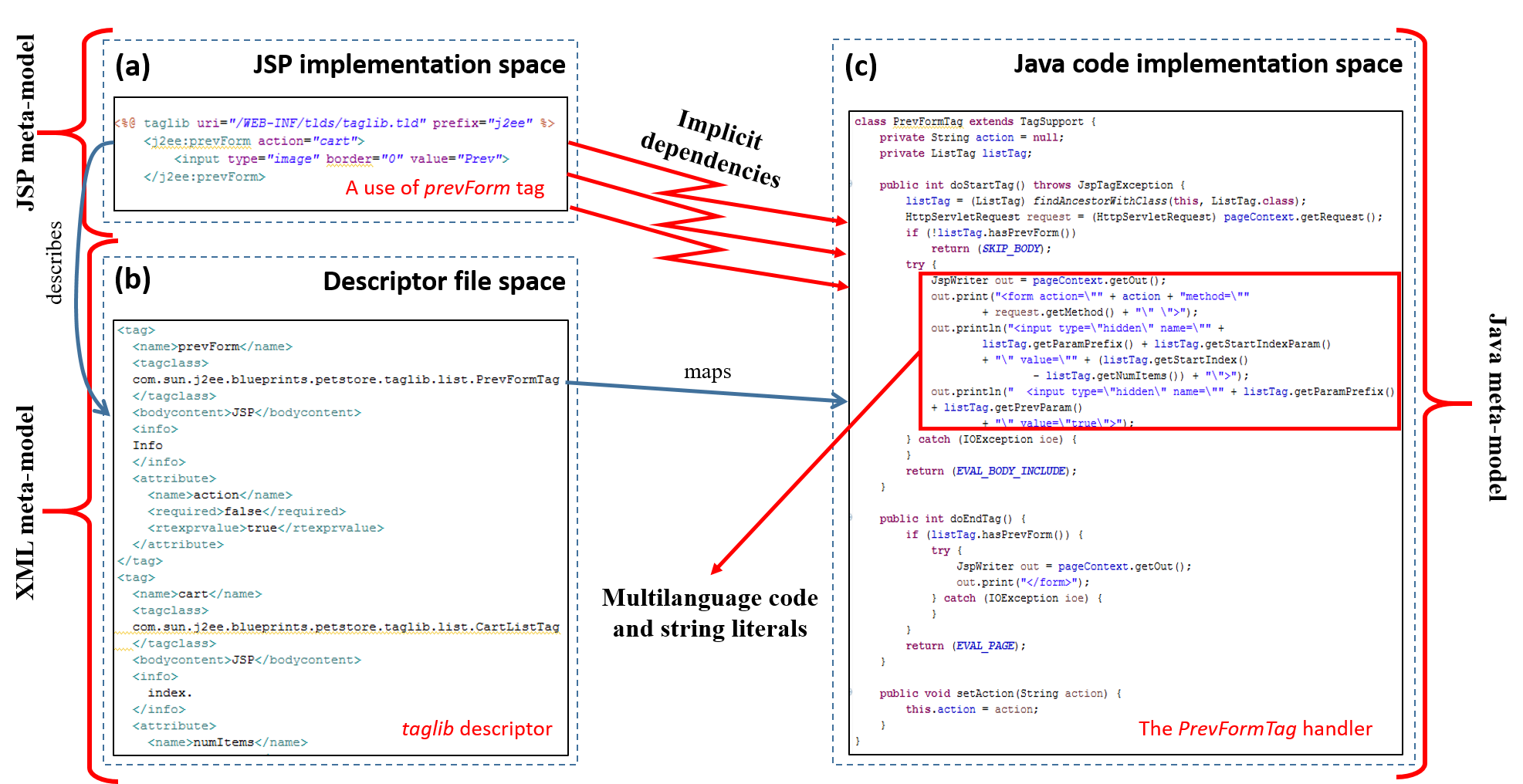}
\caption{Motivation example of JSP tag library}
\label{fig:motivation-example}
\end{center}
\end{figure*}

To better illustrate these challenges and their solutions in JEE applications, we selected the custom tag libraries from Java PetStore 1.1.2 as motivating example \cite{PetStore}. Developers of Java PetStore implemented their own \textit{custom tags} that are reused in several JSP pages. Figure \ref{fig:motivation-example}.a
%\YANN{Replace ``.a'' with a real ref using, if needed, the subfigure package} \YANN{Use proper labels, not hard-coded ones!!!} \anas{if I separate the figures I will be be able to use the lines that map dependencies between elements of different figures} \YANN{Just use subfigure and never, ever hard-code refs.} \anas{I agree with you but the lines between the sub-figures will be lost?}
shows an instance of reusing the \textit{prevForm} custom tag in a JSP page. This tag is configured using a \textit{Tag Library Descriptor} (TLD) shown in Figure \ref{fig:motivation-example}.b. TLD is an XML file that defines tags and their related attributes and also maps the tags to their implementing tag handlers, written as Java classes\footnote{A Java implementation of PrevFormTag is availabe at \url{https://docs.oracle.com/cd/E17802_01/blueprints/blueprints/code/jps11/src/com/sun/j2ee/blueprints/petstore/taglib/list/PrevFormTag.java.html}.} in Figure \ref{fig:motivation-example}.c.

\section{Multilanguage Static Code Analysis for JEE Applications}
\label{sec:change-imact-jee}

\subsection{Approach Overview}

The static analysis of JEE applications to build dependency call graphs face all the previously mentioned challenges. Figure \ref{fig:motivation-example} shows instances of these challenges (different meta-models, multilanguage code, patterns of dependencies, hidden dependencies, and string literals). Therefore, our static analysis for JEE applications must fulfill the previous requirements. 

We want to identify one dependency call graph that represents all program elements and their related dependencies in a JEE application. We must built this dependency call graph based on a language-independent meta-model that enables the representation of different meta-models (Java, JSP...) and abstracts differences among them. Also, we must represent \emph{all} dependencies that exist among program elements, regardless of their technologies (e.g., EJBs, JavaBeans, JSPs...) or application layers (e.g., Web, business).

In the literature, we identified several meta-models, such as KDM \cite{perez2011knowledge}, FAMIX \cite{famix}, Abstract Syntax Tree (AST) and 
Pattern and Abstract-level Description Language (PADL) \cite{gueheneuc2005ptidej}.
Among these meta-models, we selected the OMG KDM (Knowledge Discovery Meta-model) \cite{perez2011knowledge} to represent our dependency call graph due to:

\begin{enumerate}
\item \textbf{Possibility of evolution:} there are open-source specifications of KDM offered by the OMG that can be extended using a light-weight extension mechanism, in case we need to evolve the meta-model.

\item \textbf{Supported by open-source software tools:} KDM is supported by the \textit{MoDisco} tool \cite{bruneliere2014modisco} that offers an open-source code implementations. % Therefore, MoDisco can be extended to understand JEE dependencies.

\item \textbf{Ability to represent multilanguage software components:} KDM allows representing multilanguage software components and their related dependencies by customizing the \textit{KDMEntity} and \textit{KDMRelationship} interfaces. 

\item \textbf{Build based on the container concept:} this container concept enables the composition of an element based on other ones, e.g., a \textit{KDMEntity} instance of a class can compose a set of \textit{KDMEntity} instances corresponding to methods of this class. This supports the representation of program elements and dependencies at different levels of abstraction between (i.e., cross-language dependencies) and within (i.e., dependencies inside components) multilanguage JEE components.

\item \textbf{Ability to represent non-executable artifacts:} KDM also allows representing different configuration files that are frequently used in JEE applications by extending the \textit{KDMEntity} and \textit{KDMRelationship} interfaces.
\end{enumerate}

To build an instance of our KDM meta-model for a given JEE application, we developed a set of tools that: (1) transform JEE components into instances of KDM entities (and thei relationships), (2) describe multilanguage code in one file in the KDM model, (3) identify hidden dependencies of container services, and (4) parse string literals to identify related dependencies. We now describe in more details the technical implementation of these tools.

\subsection{Automatic Transformation of JEE Components to KDM Representations}
\label{sec.jee-to-kdm}

We developed a tool that automatically transforms the source code of a JEE application into a KDM model. First, our tool uses MoDisco to construct a KDM model of the Java program artifacts of the JEE application. Such model only\footnote{MoDisco only supports the construction of KDM models from object-oriented programming languages, like C++ and Java, not HTML or JSP} contains object-oriented elements and their related dependencies. We extended this model 
% \YANN{First, do we need to extend the meta-model with new constructs that will be instantiated?} 
% \anas{no we do not need to extend the meta-model itself at this stage.} \YANN{It's an observation or is it backed by theory? In theory, we will have to expand the meta-model... does KDM manage templates? Generics?} \anas{Ah OK, but this is another long story that we need to analyze the meta model in terms of what are the included programming languages and if it is fully or partially support them. I think we can make a feature model of programming language to help this to perform this analysis systematically? Anyways I can talk talk about it now for not open this door.}
to include program elements and dependencies of JSP and JSF pages and configuration files.

MoDisco supports the transformation of normal Java code to KDM models. Therefore, we decided to translate JSP pages to equivalent Java Servlets that are implemented by Java code because:

\begin{enumerate}
\item Servlets represent the underlying Java implementation of JSPs. Thus, they support the life cycle management of JSP pages.

\item An open-source tool is available to translate JSP pages into Servlets.

\item An open-source tool is available to build KDM models of Servlets.
\end{enumerate}

\begin{figure}[ht]
\begin{center}
\includegraphics[width=0.50\textwidth]{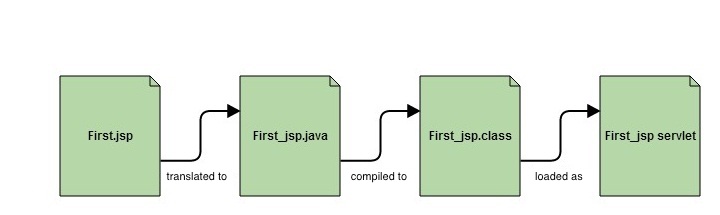}
\caption{Process of converting JSP pages into Servlets by Jasper}
\label{fig:jsp-to-servlet-translation}
\end{center}
\end{figure}

We based the translation of JSP pages to Servlets on the \textit{Jasper} tool provided by Apache Tomcat \cite{apache-tomcat}. Figure \ref{fig:jsp-to-servlet-translation} presents the process used by Jasper to translate JSP pages into Servlets.

Jasper follows a set of rules to translate each JSP page to a Servlet:

\begin{itemize}
\item \textbf{JSP scriptlet tags} (e.g., \textit{<\% code fragment \%>}) that are used to insert Java code inside the JSP pages are represented in the Servlet class with the same code fragment. For example, the \textit{scriptlet} \textit{<\% for (int i=0; i<10; i++) \%>} is translated into \textit{for (int i=0; i<10; i++)}.
	
\item \textbf{JSP declaration tags} (e.g., \textit{<\%! declaration; [ declaration; ]+ ... \%>}) that are used for variable declarations are converted into equivalent variable declarations. For example, \textit{<\%! int i=0; !\%>} is translated to \textit{int i=0;}.

\item \textbf{References to JavaBeans} are converted by creating an instance of the corresponding classes. The references to setter/getter methods are realized through normal Java method invocations. For example, \textit{<jsp:useBean id="myBeans" class= "package.BeansClass" scope="session">} is transformed into an object instantiation of \textit{package.BeansClass} and \textit{<jsp:getProperty name="myBeans" property="firstName"> </jsp:getProperty>} into a method invocation to the \textit{getFirstName()} method of this instance.

\item \textbf{Uses of custom tag handlers} are realized by a set of invocations to the life-cycle management methods of the tag handlers. Such methods are \textit{doStartTag(), doEndTag(), setAttribute()}.

\item \textbf{Other JSP tags and HTML/XML tags} are written as string literals in the output of the response objects. For example, \textit{<jsp:include page="/myPage.jsp." flush="true" />} is converted as \textit{out.write("<jsp:include page="/myPage.jsp." flush="true" />");}.
\end{itemize}

A result of these transformation rules, several, but not all, program elements and their related dependencies of JSP pages are converted to Java code. Based on this Java code, we identified an initial KDM model that includes program elements and dependencies. However, there are some JSP tags that are not translated to Java code. We identified these tags based on runtime test. We built Table \ref{table:tags-and-attributes} that summarizes the set of JSP tags and their attributes that we need to codify their dependencies. To express dependencies of these JSP tags in the initial KDM model, we developed a set of tools that: (1) parse JSP pages to identify usage scenarios of these JSP tags, (2) parse these scenarios to identify the URLs of target server pages,  (3) parse configuration files to map URLs to server pages, and (4) update our KDM model with these dependencies.

\begin{table}[ht]
\centering
\caption{JSP tags and their attributes related to dependencies}
\label{table:tags-and-attributes}
\begin{tabular}{|l|l|}
\hline
\multicolumn{1}{|c|}{\textbf{Tags}}                 & \multicolumn{1}{c|}{\textbf{Attributes}} \\ \hline
\textless form\textgreater                  & action, method                  \\ \hline
\textless jsp:include\textgreater           & page                            \\ \hline
\textless \%@ include\textgreater           & file                            \\ \hline
\textless jsp:directive.include\textgreater & file                            \\ \hline
\textless jsp:forward\textgreater           & page                            \\ \hline
\textless \%@ page \%\textgreater           & errorPage                       \\ \hline
\textless jsp:directive.page\textgreater    & errorPage                       \\ \hline
\textless a\textgreater                     & href                            \\ \hline
\textless c:redirect\textgreater            & url                             \\ \hline
\textless c:url\textgreater                 & value                           \\ \hline
\end{tabular}
\end{table}

\subsection{Describing Multilanguage Code in One File in KDM}

In JEE, multilanguage code in one file can be present in both Java classes (i.e., Servlets and tag handlers) and JSP pages.

\subsubsection{Multilanguage Code in Java Classes}

For Servlets and tag handlers, developers combine \textit{in the output stream object} (offered by the Web container) server-side Java code with a number of client-side code (i.e., \textit{$<$form action=" " ...$>$} and \textit{$<$a href$>$}). Figure \ref{fig:motivation-example}.c presents an example of an output stream object in a tag handler.

By studying Oracle's JEE specification and making runtime tests\footnote{We attached JSP tags in the output stream of a Servlet to test whether the Web container executes them at runtime and create dependencies.}, we identified that two tags are used by the Web container to describe dependencies to other service pages in the output stream: \textit{$<$form action=" " ...$>$} and  \textit{$<$a href$>$}. Therefore, we developed a tool that identifies such tags in Servlets and tag handlers implementation. It extracts URLs related to these tags based on Table \ref{table:tags-and-attributes} to identify the dependent server pages.

For example, in Figure \ref{fig:motivation-example}.c, our tool creates a dependency between the JSP page that uses this tag handler (Figure \ref{fig:motivation-example}.a) and the JSP page called \textit{cart} by analyzing the HTML form and the value of its \textit{action} attribute.

\subsubsection{Multilanguage Code in JSP Pages}

Multilanguage code can appear in JSP pages with a mix of Java code, JSP and HTML tags. We addressed such code in Section \ref{sec.jee-to-kdm} when using Jasper to transform JSP pages to Servlets. Jasper detects JSP tags (i.e., \textit{JSP declaration} and \textit{JSP scriptlet} tags) having Java code and represents their contents in the right contexts in the translated JSP pages. Also, it builds the dependencies with other JSP pages and parts thereof. For example, Jasper translates the JSP code in Listing \ref{JSPCodeExample} to the Servlet in Listing \ref{JSPinServlet}. Dependencies related to Java code are identified in the KDM using MoDisco.

\begin{lstlisting}[caption=An example JSP code, label=JSPCodeExample]
...
<TABLE BORDER="2" ALIGN="center">
<TH>Exponent</TH>
<TH>2^Exponent</TH>
<% for (int i=0; i<10; i++) {%>
<TR>
<TD><%= i%></TD>
<TD><%= Math.pow(2, i)%></TD>
</TR>
<% } //end for loop %>
</TABLE>
...
\end{lstlisting}

\begin{lstlisting}[caption=The resulting Java Servlet based on the  JSP code, label=JSPinServlet]
...
public class PowersOf2 extends HttpServlet
{
public void service(HttpServletRequest ...){
...
out.print("<TABLE BORDER='2' ALIGN='center'>");
out.print("<TH>Exponent</TH><TH>2^Exponent</TH>");
for (int i=0; i<10; i++){
out.print("<TR><TD>" + i + "</TD>");
out.print("<TD>" + Math.pow(2, i) + "</TD>");
out.print("</TR>");
} //end for loop
out.print("</TABLE>");
...
}
}
\end{lstlisting}

\subsection{Identifying hidden Dependencies of Container Services in KDM}
\label{sec:container-services}

\begin{figure}[ht]
\begin{center}
\includegraphics[width=0.47\textwidth]{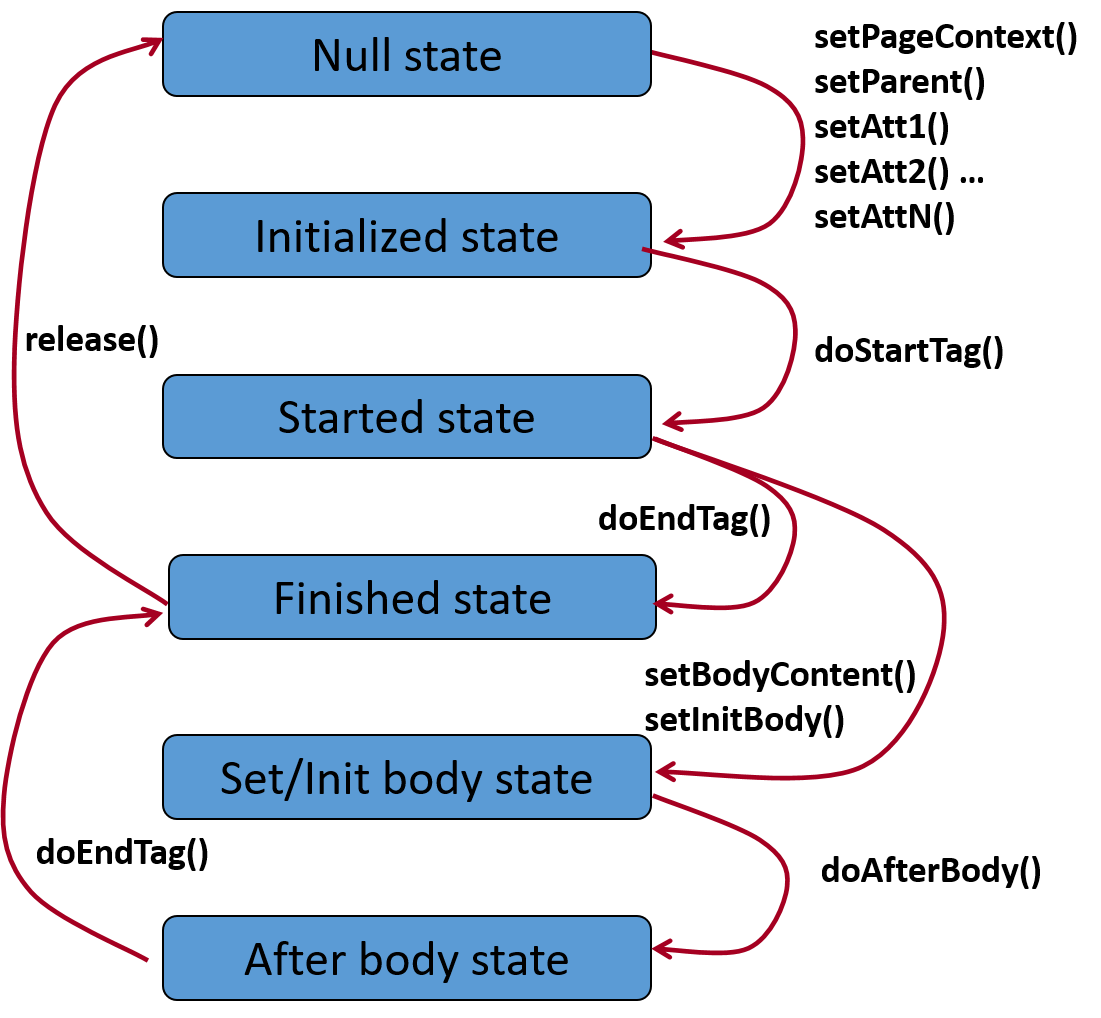}
\caption{Life cycle management of JSP tag handler}
\label{fig:life-cycle}
\end{center}
\end{figure}

JEE relies mainly on two containers to offer its services: Web and EJB containers. \textit{Web container} is for JSFs, JSPs, Servlets and tag handlers. \textit{EJB container} is for Enterprise Beans. We studied these two containers and identified a set of patterns that they use to offer their services (e.g., life cycle management, RMI, callback methods, etc.). We built state diagram models based on the life cycles of JEE components. Figure \ref{fig:life-cycle} shows an instance of these state diagrams related to EJBs and JSP tag handlers.

We observed that some dependencies are configurable, like security, transaction, and persistence. Their configuration parameters are specified in XML configuration files (e.g., \textit{web.xml}, \textit{ejb.xml}, \textit{tag library descriptor.tld}) and code annotations (\textit{web annotations}, \textit{EJB annotations}). We studied patterns/idioms that JEE containers use to parse these configuration files and developed a specific parser for each configuration file to extract dependencies.  

For our example in Figure \ref{fig:motivation-example}.b, our parser identifies that \textit{prevForm} has a mandatory attribute called \textit{action} handled by the \textit{PrevFormTag} class located in the \textit{com.sun.j2ee.blueprints.petstore.taglib.list} package.

\begin{figure*}[ht]
\begin{center}
\includegraphics[width=\textwidth]{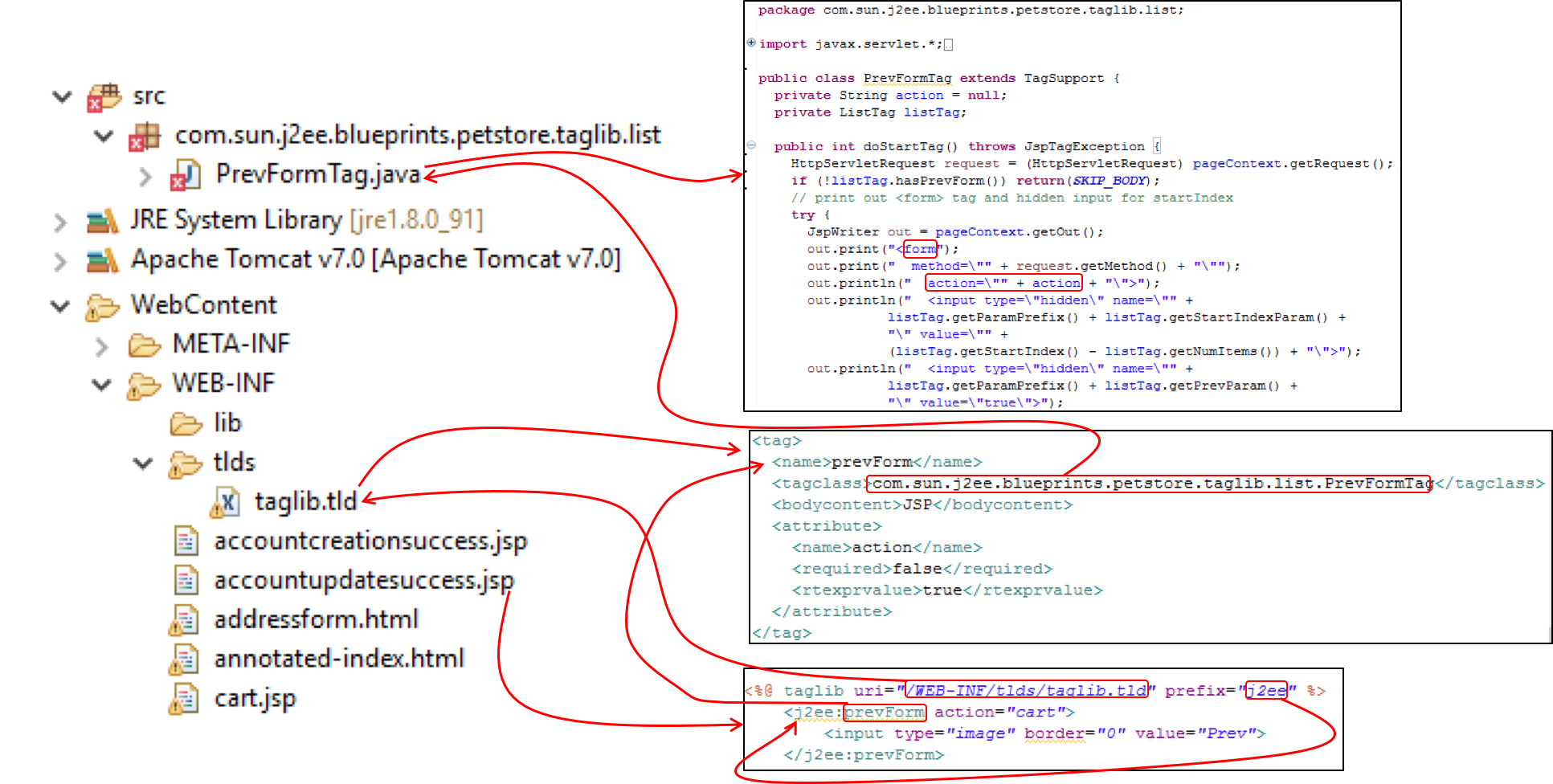}
\caption{Container service dependencies of our motivating example in Figure \ref{fig:motivation-example}}
\label{fig:example-container-service-dependeices}
\end{center}
\end{figure*}

Figure \ref{fig:example-container-service-dependeices} summarizes the container service dependencies that our tool identifies when a JSP page uses prevForm. These dependencies are due to callback methods: \textit{setAction("cart")} as setter method for the \textit{action} attribute and \textit{doStartTag()} and \textit{doEndTag()} for life cycle management of the tag handler. In our KDM model, we created dependencies from the KDM entity of this JSP page to the KDM entities of the corresponding methods of \textit{PrevFormTag}. We proceeded  similarly for other JEE components; Servlets, JSP pages, ejb.

\subsection{Parsing String Literals to Identify Dependencies}

\begin{figure}[ht]
\begin{center}
\includegraphics[width=0.5\textwidth]{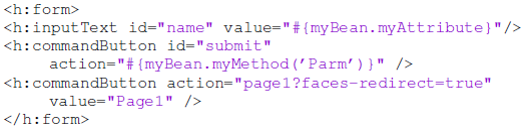}
\caption{String literals in server pages}
\label{fig:string-literals-server-pages}
\end{center}
\end{figure}

String literals are used for multiple purposes. Figure \ref{fig:string-literals-server-pages} shows string literals used to forward requests between server pages and to communicate with Beans components, e.g., access attributes (the value entered by the end-user is stored in \textit{myBeans.myAttribute}) and method invocations (the value returned from \textit{myBeans.myMethod()} determines the relative-URL). We identified dependencies in string literals as follows: 

\begin{enumerate}
\item We analyzed the JEE specification and observed the situations where string literals create dependencies.

\item We built a list of tags and their related attributes in which string literals codify dependencies.

\item We developed a string parser that analyzes the values corresponding to the tags and attributes in our list. This parser is based on \textit{Expression Language} because it is the official language used by the JSP engine to construct such string literals.
\end{enumerate}

\section{Evaluation}
\label{sec:evaluation}

We now evaluate our static code analysis approach for JEE applications. We developed DeJEE (\textbf{De}pendencies in \textbf{JEE}), a fully automated tool to identify a dependency call graph of a given JEE application based on the KDM meta-model. 
We tested DeJEE on two case studies: Java PetStore \cite{PetStore} and JSP Blog \cite{JSPBlog}. 

We selected Java PetStore because it represents the official demonstration example provided by Sun Microsystems on how to develop flexible, scalable, cross-platform JEE applications. Also, the availability of its source code and documentation makes it ideal. Finally, it is implemented using many JEE technologies (e.g., JSPs, Servlets, EJBs, etc.). This implementation is composed of 88 JSP, 233 Java classes, and 8 HTML files. 

JSP Blog is implemented using 10 JSP and 1 HTML files. We selected JSP Blog due to the fact that its JSP pages intensively use multilanguage code (combination of JSP tags and Java code) and do not depend on Java classes, which allows us to expose the limitations of existing tools, i.e., Modisco. Thus, it is a good option for evaluating how DeJEE works with multilanguage files but without Java classes.

\subsection{Objectives and Methodology}

We want to achieve the following three objectives:

\begin{enumerate}
\item \textit{\textbf{Objective:}} Demonstrate the capability of DeJEE to meet the five requirements of static code analysis in the case of JEE applications. 

\textit{\textbf{Method:}} we extract one common KDM model for each case study. This addresses the first challenge related to dealing with diversity of meta-model. Then, we highlight the identified dependencies related to the other four challenges of analyzing multilanguage code in one file, patterns of dependencies, hidden dependencies on container services and dependencies encoded in string literals.

\item \textit{\textbf{Objective:}} Measure the improvement in recall (if any) that our approach adds in comparison to MoDisco in extracting multilanguage dependencies.

\textit{\textbf{Method:}} we extract KDM models using both MoDisco and DeJEE and then we compare the KDM models to report their differences in terms of numbers of KDM entities and relationships, which describe dependencies among JEE program elements. We selected MoDisco because it is the only tool that is remotely comparable to DeJEE in terms of their KDM models.

\item \textit{\textbf{Objective:}} measure the correctness of dependencies identified by DeJEE.

\textit{\textbf{Method:}} 
we manually constructed a ground-truth call graph of JSP Blog  to precisely measure the precision and recall of dependencies identified by DeJEE. We identified the list of JSP pages invoked by each JSP page. 

As it is hard to manually construct a ground-truth call graph of Java PetStore due to its large number of source code files (88 JSP, 233 Java classes, and 8 HTML files.), we only manually evaluated the precision of identified dependencies in Java PetStore. For an identified KDM relationship instance, we studied the implementation files of the two related KDM entities to validate whether these implementation files do have a dependency or not.
\end{enumerate}

\subsection{Results}

\subsubsection{Capability of DeJEE to Identify Dependencies}

Table \ref{table:dependencies-results} shows the results of applying DeJEE on Java PetStore and JSP Blog.

%For instance, we manually checked multilanguage files these files and find that 100\% of them contained multilanguage code. 

\paragraph{Multilanguage Code in One File} The results show the ability of DeJEE to automatically detect multilanguage files. DeJEE identified 40 and 6 files respectively in Java PetStore and JSP Blog. We manually checked these files and found that 100\% of them contained multilanguage code. 
By analyzing these files, we found that DeJEE identifies dependencies related to multilanguage code of Servlets, tag handlers and JSP pages in the KDM model. 
For JSP Blog, DeJEE discovers 41 and 20 program dependencies by analyzing Java code and JSP tags in the multilanguage files, respectively.

\paragraph{Pattern of Dependencies} Different communication patterns are used to create dependencies among different JEE components (JSPs, JavaBeans, EJBs, Managed Beans, etc.). The \textit{HTTP Request Communication Pattern} is used by JEE components to invoke Servlets and JSP pages. JSPs and JSFs use special tags and an expression language to connect to JavaBeans and Managed Beans components. EJBs are invoked through a \textit{JNDI} lookup (Java Naming and Directory Interface), e.g., \textit{context.lookup("java:comp/env/ejb/HelloBean");}. JSPs use this interface as Java code embed using the \textit{scriptlet} tag. DeJEE can recover these dependencies using these patterns.

\paragraph{Container Service Dependencies} DeJEE identified container service dependencies related to \textit{custom taglib}, JSPs, and Servlets by parsing related configuration files and annotations. To map relative-url to their corresponding Servlets or JSP pages, we parse \textit{web.xml} and web annotations. The mapping of custom taglib to identify tag handlers is based on \textit{taglib.tld}.

\paragraph{Dependencies in String Literals} DeJEE discovered string literals and parsed them to identify embedded, hidden dependencies. On average, each JSP page contains 17.51 (1,541/88 pages) and 18.54 (204/11 pages) string literals in which 46.7\% (720/1,541) and 31.8\% (65/204) of these string literals are related to dependencies in PetStore and JSP Blog, respectively.

In summary, DeJEE successfully identified 329 and 11 \textit{KDMEntity} instances corresponding to Java code, JSP, and HTML files, and 2,673 and 61 \textit{KDMRelationship} instances corresponding to dependencies, respectively for PetStore and JSP Blog.

\begin{table}[ht]
\centering
\caption{Results of DeJEE tool}
\label{table:dependencies-results}
\begin{tabular}{lcc}
\hline
\multicolumn{1}{|c|}{Results}                            & \multicolumn{1}{c|}{Java PetStore}  & \multicolumn{1}{c|}{JSP Blog}       \\ \hline
\multicolumn{1}{|l|}{Multilanguage files}                & \multicolumn{1}{c|}{45.4\% of JSPs} & \multicolumn{1}{c|}{54.5\% of JSPs} \\ \hline
\multicolumn{1}{|l|}{Total no.\ of string  literals}      & \multicolumn{1}{c|}{1541}           & \multicolumn{1}{c|}{204}            \\ \hline
\multicolumn{1}{|l|}{No.\ of string literals having dep.} & \multicolumn{1}{c|}{720}            & \multicolumn{1}{c|}{65}             \\ \hline
\multicolumn{1}{|l|}{No.\ of KDM entities}                & \multicolumn{1}{c|}{329}            & \multicolumn{1}{c|}{11}             \\ \hline
\multicolumn{1}{|l|}{No.\ of KDM relationships}           & \multicolumn{1}{c|}{2673}            & \multicolumn{1}{c|}{61}             \\ \hline              
\end{tabular}
\end{table}

\subsubsection{Recall Improvement}

\begin{figure}[ht]
	\begin{center}
		\includegraphics[width=0.5\textwidth]{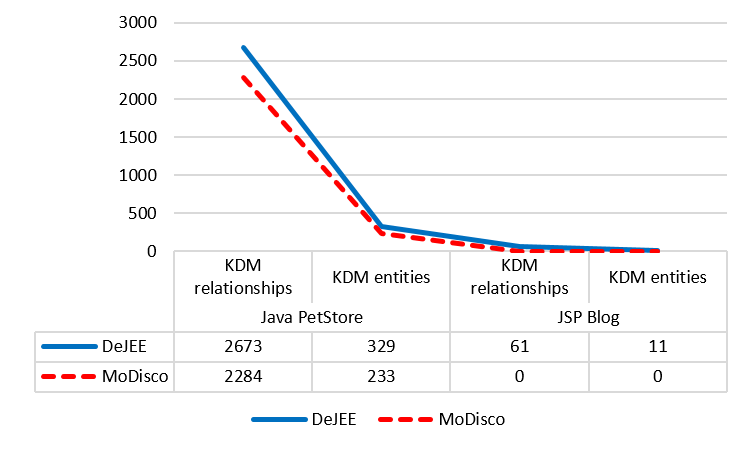}
		\caption{Recall Improvement}
		\label{fig:recall-resutls}
        \end{center}
\end{figure}

Figure \ref{fig:recall-resutls} shows the results of the recall improvement that DeJEE achieves compared to MoDisco. MoDisco does not support the analysis of JSP pages and, thus, cannot build KDM models for JSP pages and identify dependencies in these pages or between these pages and other program elements, like Java code. Also, when analysing multilanguage code in one file, MoDisco outputs an empty KDM model for JSP Blog while DeJEE provides a KDM model that consists of 11 \textit{KDMEntity} and 61 \textit{KDMRelationship} instances. Similarly for Java PetStore, DeJEE provides 329 \textit{KDMEntity} and 2673 \textit{KDMRelationship} instances compared to 233 \textit{KDMEntity} and 2284 \textit{KDMRelationship} ones provided by MoDisco.

Consequently, DeJEE improves MoDisco's recall by discovering 41\% ($(329-233)/233$) and 100\% ($(11-0)/11$) more program elements and 17\% ($(2,673-2,284)/2,284$) and 100\% ($(61-0)/61$) more program dependencies, respectively in Java PetStore and JSP Blog, than MoDisco. DeJEE provides better results because of its codification of dependencies related to container services, string literals, and multilanguage code.

\subsubsection{Correctness of Dependencies Identified by DeJEE}
For JSP Blog, we found that the constructed ground truth call graph includes invocations that confirm to the KDM model identified by DeJEE with 100\% precision and 100\% recall. 

It is worth to recall that we only measure the precision of the KDM dependencies of Java PetStore due to difficulties in building ground truth call graph of Java PetStore (i.e., size of system).
We found that the identified KDM dependencies of the Java PetStore KDM model are true with 100\% precision when we manually evaluated them. 

% \subsection{Discussions}
% \anas{Only one subsection still!}
% \subsubsection{Different Implementation Patterns}
% In our case studies, we identified two patterns that are implemented in different ways to map HTTP requests in addition to the standard one implementing using the \textit{web.xml} file. These patterns are \textit{Front Controller Pattern} and \textit{Direct Call Pattern} ones.
% Java PetStore relies on the \textit{Front Controller Pattern} such that the \textit{MainServlet.java} Servlet processes a part of requests arrived to the website \cite{nambiar:petstore}. Then, \textit{MainServlet.java} determines/invokes the corresponding Servlets and JSP pages. For this case, we studied the semantic of \textit{MainServlet.java} to develop an automated parser to identify url mapping to Servlets and JSPs. 
% In the other hand, JSP Blog implements the \textit{Direct Call Pattern}. JSP pages call each other directly through their names. 

\subsection{Threats to Validity}
In this section, we discuss internal and external threats to validity of the proposed approach.

\subsubsection{Internal Threads to Validity}

\paragraph*{Manual Construction of Ground Truth Call Graphs}
We conducted the evaluation of the resulting KDM models by manually constructing a ground-truth dependency call graph of JSP Blog. We report that this manual construction could be error prone in which any error will affect the precision and the recall values. To avoid such error prone, we validated the constructed call graph three times before we started the comparison with the DeJEE's ones. Note that we are familiar with this case study since we are working with it two years ago as a part of our project that aims to analyze program dependencies in JEE Web applications.   

We plan to measure the precision and recall of dependencies identified by DeJEE compared to a real ground truth dependency call graphs. We want to compare DeJEE dependencies against ones obtained from runtime executions.
In this context, dependencies can be classified into three types as follows: (i) dependencies that appear in both cases have 100\% precision.
(ii) Dependencies that are only identified during the runtime execution are considered as references to identify true negative dependencies that DeJEE does not identify.
(iii) Dependencies that are only identified by DeJEE are under thought as the runtime execution may not cover their execution paths following the tested use case scenarios of the software. In this case, we will rely on external human experts to validate the correctness of DeJEE. We will try to get in touch with developers of case studies to evaluate them. Otherwise, we will consider M.Sc.\ and Ph.D.\ students.

\paragraph*{Missing of Java Reflection Dependencies} We are aware that there are missing dependencies in DeJEE dependency call graphs duo to Java reflection. We want to address this limitation in future work to improve further the recall of DeJEE. We will rely on the study of Landman et al. \cite{landman2017challenges} and Livshits et al. \cite{livshits2005reflection} to solve Java reflexion challenges.

\subsubsection{External Threads to Validity}

\paragraph*{Coverage of DeJEE for Other Case Studies} When we developed DeJEE, we covered as many dependency codification patterns, which can be used by developers, as possible. To do so, we studied the Oracle's JEE specification, which is the main reference for JEE developers to learn the JEE technologies. For example, we identified the various mechanisms that can be used in JSP pages to call another ones. This represents 100\% of pure inter-JSP dependencies.
We applied DeJEE on two JEE applications; Java PetStore and JSP Blog. As we mentioned earlier, Java PetStore is the official demonstration example provided by Sun Microsystems to explain how to develop JEE applications using several implementation patterns. JSP Blog can be considered as a good representative of JSP pages that mix Java code due to the fact that its JSP pages intensively use multilanguage code (combination of JSP tags and Java code). All of these argue that DeJEE can work for most of JEE applications that follow standard JEE development patterns.

\paragraph{Including Dependencies of Other Programming Languages}
DeJEE builds dependency call graphs based on the KDM meta-models that supports the representation of other programing languages. Therefore, we are allowed to extend DeJEE to include other programming languages (e.g., C++). To do this extension, we have to: 
(i) transform source code elements of these programming languages to KDM representations, (ii) identify patterns of dependencies in the new multilanguage code (e.g., extract JNI dependencies), (iii) develop a tool that detects these patterns, and (iv) extend the KDM model by adding identified dependencies.

% \textbf{Comments form SANER: We want to show in this section that the selection and processing of the data have not affected the final results. Also, it is important to clarify how could the results be generalized to other multi-language software with a different set of languages. This would make it easier for other researchers to understand the limitations of this work and hence make improvements if possible.}

% \subsection{Examples of DeJEE Use Cases}

% We discuss some example use cases of DeJEE to demonstrate how software engineers (i.e., the actor) can get benefits from it. \YANN{This section seems quite empty... Why all these subsubsections? Why are there no explanations with each use case, at least on how DeJEE could help?}

% \subsubsection*{Use Case 1: Identify what JEE components will be impacted when I make a change in a JSP page.}

% \subsubsection*{Use Case 2: Identify what JEE components have multilanguage code in a JEE application.}

% \subsubsection*{Use Case 3: Parse configuration files to identify container--service dependencies.}

% \subsubsection*{Use Case 4: Discover dependencies among program elements that do not exist explicitly in the source code.} 

% \subsubsection*{Use case 5: Identify what is the program slice of a given program element.}

% \subsubsection*{Use Case 6: Discover dependencies among program elements that do not exist explicitly in the source code but would cause runtime errors if missing.}

% \subsubsection*{service identification} 

\section{Related Work}
\label{sec:related-work}
% Existing traditional \emph{unilingual} static code analysis approaches must be evolved by considering other kinds of analyses, involving other kinds of artifacts and codifying container services provided by frameworks. We explain in Figure \ref{fig:dependency-call-graphgs} the idea of improving the recall of static code analysis approaches.

Although many research works have been presented in the literature to support the static code analysis of multilanguage systems \cite{von1995program,kienle2010rigi,bruneliere2014modisco,moise2005extracting,kraft2008cross,german2009license}, these works suffer from three main limitations.
First, they are able to analyze one programming language at one time, although they can deal with more than one programming language in isolation from one another like \cite{von1995program,bruneliere2014modisco,kienle2010rigi}. Second, they do not identify hidden dependencies related to container services \cite{bruneliere2014modisco,moise2005extracting,kraft2008cross,german2009license}. Third, they are not able to analyze source code files mixing multilanguage code \cite{von1995program,kienle2010rigi,bruneliere2014modisco,moise2005extracting}. Furthermore, none of these existing works have described the challenges---and their requirements---in the static code analysis of multilanguage software systems. Instead, each focused on providing solutions of a narrow, particular set of challenges.

Mayrhauser and Vans \cite{von1995program} studied various program comprehension models and sketched an integrated model that can be used to explain developers' comprehension of components written in any programming language one at a time. As another example, Muller and his group developed Rigi \cite{kienle2010rigi}, an environment to reverse engineer, explore, visualize, and document components in C, C++, or COBOL; in isolation from one another.

Cross-language research works are intrinsically multilanguage and their results pertain to the dependencies among heterogeneous components, two or more at a time. Such dependencies include (1) procedure calls; (2) clones; (3) license integration patterns; and (4) refactorings. 
For example, Moise and Wong \cite{moise2005extracting} were among the first researchers, e.g., \cite{linos2003tool, deruelle2001analysis} to extract, represent, and study cross-language dependencies, i.e., among heterogeneous components. They used the API provided by each language to identify cross-language calls, e.g., calls to Java Native Interface API in C and Java. 
Kraft et al. \cite{kraft2008cross} developed a technique to identify cross-language clones using the Microsoft Code-DOM library for .NET languages and a hybrid token/tree-based algorithm for clone detection. They reported clones whose siblings exist in components written in both C\# and Visual Basic.NET. 
German and Hassan \cite{german2009license} described five possible kinds of dependencies between heterogeneous components. These are linking, forking, sub-classing, inter-process communication, and plug-in. Then, they identified these kinds of dependencies in 124 open-source software systems. Using the identified component dependencies and their licenses, they proposed 12 patterns of license integrations.
Mayer and Schroeder \cite{mayer2012cross} proposed a technique based on the MOF Query, View, Transformation Relations specification (QVT/R) to identify dependencies among heterogeneous components, warn of potential missing dependencies, and propagate renamings among heterogeneous components.

Although of major importance to understand multilanguage software, these works do not analyze the heterogeneous components as wholes, i.e., the patterns of dependencies and patterns of control and data flows through these dependencies. Only few works focused explicitly on the interactions among heterogeneous components, taken together: (1) static and (2) dynamic data and control flow interactions and (3) conformance of components and their configurations.
Ayers et al. \cite{ayers2005traceback} proposed TraceBack to diagnose bugs in multilanguage software by collecting data through runtime instrumentation of control-flow blocks. The data is collected by statically rewriting the binaries and--or instrumenting the intermediate languages to generate a unified trace of the components' execution.
Tan and Croft \cite{tan2008empirical} studied the interactions between Java code and the C++ code in the underlying Java virtual machine. They showed that bugs are possible due to the assumptions made by the Java code regarding the C++ code. For example, the C++ native method \textit{java.util.zip.Deflater.deflatesByte()} assumes that its Java callers check bounds, which could lead to buffer overflows. 

Some works also relied on KDM models to analyze multilanguage code, 
Yazdanshenas and Moonen \cite{yazdanshenas2011crossing} built homogeneous KDM models of heterogeneous systems, with components in C, C++, and Java and configuration files in XML. They used these models to obtain system dependency graphs and sliced these graphs to show if a given input is used to produce the expected output, typically in sensor/actuator systems and other such component-based systems. However, their work only takes into account object-oriented meta-model programming languages (similar to what MoDisco does \cite{bruneliere2014modisco}), disregarding other non object-oriented languages like JSPs and JSFs. 

For the case of Web applications, Ricca et al. \cite{ricca2002construction} sliced Web applications based on a dynamic analysis techniques. They traced dependencies between the PHP server and the Javascript client to extract the program elements relevant to a specified computation. Naumovich et al. \cite{naumovich2004static} proposed a static analysis approach only for EJBs and JEE access policies. Kirkegaard et al. only analyzed output streams of Servlets to verify if they confirm the container specifications using context free grammars. 
Perin \cite{perin2012reverse} introduced a meta-model and revere-engineering techniques to model the components of multilanguage software, focusing on Enterprise Java Beans software. Then, he used his approach to validate their architectures, to map their databases and transactions flows, and to identify hidden (i.e., abstracting explicit) dependencies as well as visualization techniques to ease their understanding. Although these approaches identified dependencies that are similar to ones in JEE applications, they are ad-hoc and did not cover all types of JEE dependencies. Shatnawi et al.  \cite{shatnawi2017analyzing} \cite{shatnawi2018identifying} identified KDM model of JEE applications by analyzing Servlets and JSPs. However, they did not analyze dependencies related to tag handler dependencies. Hecht et al. \cite{hecht2018codifying} extracted a declarative specification of the hidden dependencies of the J2EE applications that are inherent in the services offered and that are not visible in the user code that uses them. Based on this declarative specification, they codified of the hidden dependencies into rules that can be automatically detected using a tool. 
Despite the large contributions of these works in analyzing program dependencies in multilanguage software in general and JEE applications in specific, none of them describe the main challenges--and their consequence requirements--in analyzing multilanguage software systems. Also, existing works do not cover all dependencies that are resulted in multilanguage JEE application components. For example none of these works allowed us to codify the dependencies in our motivation example in Figure \ref{fig:motivation-example}.

\section{Conclusion and Future Work}
\label{sec:conclusion}

%\subsubsection*{Conclusion}
The static code analysis is difficult when dealing with multilanguage software systems, such as JEE applications, which combine Java, JSF, JSP, and other source code and configuration files. In this paper, we identified five challenges with multilanguage software systems and proposed requirements to address them when developing static code analysis approach of JEE applications. 

We developed an automated analysis tool, DeJEE, that illustrates our approach to identify a dependency call graph of a given JEE application. This dependency call graph is identified based on the KDM meta-model. We developed DeJEE as Eclipse-plug-in that allows software developers: (i) to identify files having multilanguage code in JEE applications, (ii) to describe program elements and their dependencies across and within Java code, Servlets, and JSPs using one common KDM model, (iii) to parse configuration files to identify container--service dependencies, and (iv) to discover dependencies among program elements that do not exist explicitly in the source code of the analyzed applications and would cause runtime errors if missing.

We applied DeJEE on two JEE applications (Java PetStore  and JSP Blog) to evaluate the capability of DeJEE to solves the challenges of static code analysis in JEE applications and the correctness of DeJEE identified dependencies. Our results showed that DeJEE solves the challenges that we identified in the static analysis of JEE applications and detects related dependencies with 100\% precision for JSP Blog and Java PetStore, and with 100\% recall for JSP Blog. 
Furthermore, we compared DeJEE with MoDisco and found that DeJEE improves the recall of MoDisco by discovering respectively 70.5\% and 58.5\% more KDM entity instances and KDM relationship instances for the two case studies.

%\subsubsection*{Future Work}
We plan to extend our work in the following four directions.

\begin{enumerate}
\item \textit{Generalization of DeJEE:} we plan to generalize the scalability DeJEE by experimenting it on a number of JEE applications of different sizes and from various domains.

\item \textit{Extension of DeJEE:} we want to extend DeJEE by addressing static analysis of Java reflexion. We will rely on study of Landman et al. \cite{landman2017challenges} to solve Java reflexion challenges which is used in Web applications. Also, we plan to extend DeJEE to analyze dependencies in multilanguage software including other programming languages such as JavaScript, PHP.

\item \textit{Application of DeJEE:} we want to exploit the KDM model identified by DeJEE to perform other software engineering tasks like program understanding, change impact analysis, program debugging,  reverse engineering and service identification. 
%To do so, we will extract a multi-level dependency call graph from the KDM model. This dependency call graph should be able to represent program elements and their related dependencies corresponding to multilanguage software at different levels of abstraction. Such levels are related to components, files, functions and statements.

\item \textit{Study the Prevalence of the Identified Challenges in Other Multilanguage Software Systems:} We identified, in this paper, five main challenges related to the static analysis of multilanguage software. We want to empirically study how frequently these challenges occur in other multilanguage software systems (rather than JEE applications), which supports the extension of DeJEE.
\end{enumerate}

\bibliographystyle{unsrt}
\bibliography{sigproc}
\end{document}